\documentclass[aps,prc,twocolumn,floatfix,showpacs,amsmath,amssymb]{revtex4}
\usepackage{graphicx}
\usepackage{dcolumn}
\usepackage{color} 
\begin{document}
\preprint{UCRL-JRNL-217508}
\title{Proton radii of $^{4,6,8}$He isotopes from high-precision nucleon-nucleon interactions} 
\author{E. Caurier}
\email[]{etienne.caurier@ires.in2p3.fr}
\affiliation{$^3$Institut de Recherches Subatomiques
            (IN2P3-CNRS-Universit\'e Louis Pasteur)\\
            Batiment 27/1,
            67037 Strasbourg Cedex 2, France}
\author{P. Navr\'atil}
\email[]{navratil1@llnl.gov}
\affiliation{Lawrence Livermore National Laboratory, P.O. Box 808, L-414,
Livermore, CA  94551, USA}

\date{\today}

\begin{abstract}
Recently, precision laser spectroscopy on $^6$He atoms determined accurately
the isotope shift between $^4$He and $^6$He and, consequently, the charge radius 
of $^6$He. A similar experiment for $^8$He is under way. We have performed 
large-scale {\it ab initio} calculations for $^{4,6,8}$He isotopes using
high-precision nucleon-nucleon (NN) interactions within the no-core shell model
(NCSM) approach. With the CD-Bonn 2000 NN potential we found point-proton 
root-mean-square (rms)
radii of $^4$He and $^6$He 1.45(1) fm and 1.89(4), respectively, 
in agreement with experiment
and predict the $^8$He point proton rms radius to be 1.88(6) fm. At the same time, 
our calculations show that the recently developed nonlocal INOY NN potential 
gives binding energies closer to experiment, but underestimates the charge radii.
\end{abstract}

\maketitle
%

Recent advances in the theory of the atomic structure of helium \cite{atomic_str}
as well as in the techniques of isotopic shift measurement made it possible
to determine accurately the charge radius of $^6$He \cite{6He_chrad}. Precision
laser spectroscopy on individual $^6$He atoms confined and cooled in a 
magneto-optical trap was performed and measured the isotope shift between
$^6$He and $^4$He. With the help of precise quantum mechanical calculations
with relativistic and QED corrections \cite{Drake} and from the knowledge of 
the charge radius of $^4$He (1.673(1) \cite{4He_chrad}), 
it was possible to determine the charge radius of $^6$He to be 
$2.054\pm0.014$ fm \cite{6He_chrad}. The large difference between the $^4$He 
and $^6$He charge radii is due to the extra two loosely bound neutrons in $^6$He 
that form a halo \cite{Tanihata}. A similar experiment to determine the
charge radius of $^8$He is under way \cite{8He_chrad}.

It is a challenge for {\it ab initio} many-body methods to calculate 
the nuclear radii with an accuracy comparable to current experimental 
accuracy and test in this way the nuclear Hamiltonians used as 
the input of {\it ab initio} calculations. At present, there are two
{\it ab initio} approaches capable of describing simultaneously 
the $^4$He, $^6$He and $^8$He isotopes starting from realistic 
inter-nucleon interactions. One is the Green's 
function Monte Carlo (GFMC) method \cite{GFMC_04} and the other is the 
{\it ab initio} no-core shell model (NCSM) \cite{NCSMC12}. In this paper,
we calculate the ground-state properties of $^4$He, $^6$He and $^8$He 
within the NCSM. We test two vastly different accurate nucleon-nucleon (NN) 
potentials, the CD-Bonn \cite{cdb2k} and the the INOY
(Inside Nonlocal Outside Yukawa)~\cite{dol03:67,dol04:69}.

In the NCSM, we consider a system of $A$ point-like non-relativistic
nucleons that interact by realistic two- or two- plus three-nucleon
interactions. The calculations are performed using a finite harmonic
oscillator (HO) basis. As in the present application we aim at
describing loosely bound states, it is desirable to include as many
terms as possible in the expansion of the total wave function. By
restricting our study to two-nucleon (NN) interactions, even though
the NCSM allows for the inclusion of three-body forces \cite{v3b},
we are able to maximize the model space and to better observe the
convergence of our results. The NCSM theory was outlined in many
papers. Here we only repeat the main points.

We start from the intrinsic two-body Hamiltonian for the $A$-nucleon system
$H_A=T_{rel} + {\cal V}$, where $T_{rel}$ is the relative kinetic energy
and ${\cal V}$ is the sum of two-body nuclear and Coulomb
interactions. Since we solve the many-body problem
in a finite HO basis space,
it is necessary that we derive a model-space dependent effective
Hamiltonian. For this purpose, we perform
a unitary transformation \cite{NCSMC12,NKB00,LS81,UMOA}
of the Hamiltonian, which accommodates the short-range correlations.
In general, the transformed Hamiltonian is an $A$-body operator.
Our simplest, yet non-trivial, approximation that we employ in this work
is to develop a two-particle cluster effective Hamiltonian, while
the next improvement is to include three-particle clusters, and so on.
The effective interaction is then obtained
from the decoupling condition between the model space and the excluded space
for the two-nucleon transformed Hamiltonian.
The resulting two-body effective Hamiltonian
depends on the nucleon number $A$, the HO frequency $\Omega$, and
$N_{\rm max}$, the maximum many-body HO excitation energy
defining the model space.
It follows that the effective interaction, which is translationally 
invariant, approaches the starting bare
interaction for $N_{\rm max}\rightarrow \infty$. Consequently, by construction
the method is convergent to the exact solution.
At the same time, the NCSM effective interaction method is not variational 
as higher-order terms may contribute with either sign to total binding.

Once the effective interaction
is derived, we diagonalize the effective Hamiltonian
in a Slater determinant (SD) HO basis that spans a complete
$N_{\rm max}\hbar\Omega$ space. We have reached model spaces
of $N_{\rm max}=22,16$ and 12 for $^4$He, $^6$He and $^8$He, 
respectively. This is a highly non-trivial problem. The dimensions
are large, e.g. $7\times10^8$ for $^6$He, although still smaller
than in standard shell model calculations, e.g. the dimension
is $10^9$ for $^{56}$Ni in full $fp$-shell. 
The first difficulty is due to the large number of shells. 
In the $N_{\rm max}=22$ model space, there are 276 $nlj$-shells 
corresponding to 4600 $nljm$ individual states. This can be compared
4 and 20, respectively for $^{56}$Ni in full $fp$-shell. 
This means that one has to handle a huge number of operators. 
Therefore, it has been necessary to write a specialized version of the shell 
model code Antoine~\cite{EC99,Antoine}, suitable for the NCSM applications,
see, e.g., Refs. \cite{Be8,A10_NCSM}.
The code works in the $M$ scheme for basis states, 
and uses the Lanczos algorithm for diagonalization.
Its basic idea is to write the basis
states as a product of two Slater determinants, a proton one 
and a neutron one.
Matrix elements of operators are calculated for each separate subspace
(one-body for the proton-neutron, two-body for the proton-proton 
and neutron-neutron).
The performance of the code is the best when the ratio between the number
of proton plus neutron SD and the dimension of the matrix is the least. 
It happens when the number of proton SD is equal the number of the neutron SD.
To highlight the differences between the standard and the NCSM calculations, 
we note that for example, in $^{56}$Ni there are 125970 neutron SD, while 
in $^6$He and $^8$He there are $19.5\times10^6$ and $43\times10^6$ neutron 
SD, respectively. Further, for example, in the basis of $^{48}$Ca
the 12022 neutron SD  with $M=0$ produce 144528484 states
in the full basis. In $^8$He, the 8986408 neutron SD with $M=0$ produce
only 56216057 states in the full basis and 7007190 of these SD that have
$N=12$ are associated with a unique $N=0$ proton SD.
Another comment concerning the difficulty of performing the NCSM calculations 
is that the matrices become less sparse when the number of particles decrease.
To keep the comparison with the standard shell model, in $^4$He we have 
a dimension 12.5 smaller than in $^{56}$Ni but 1.5  times 
more non-zero matrix elements.
For all these reasons, NCSM calculations with large $N_{\rm max}$ model spaces
are difficult but still feasible with a computer with a large RAM memory 
and a large disk capacity. As a last example, one Lanczos iteration in $^6$He 
takes 7 hours while the same in $^{56}$Ni takes 70 minutes on an Opteron machine.

As already mentioned, we test two different, high-precision NN interactions 
in this study: the CD-Bonn 2000 ~\cite{cdb2k} and the 
INOY~\cite{dol03:67,dol04:69} potentials.

The CD-Bonn 2000 potential \cite{cdb2k} as well as its earlier version \cite{cdb}
is a charge-dependent NN interaction based
on one-boson exchange. It is described in terms of covariant Feynman
amplitudes, which are non-local. Consequently, the off-shell behavior of
the CD-Bonn interaction differs from local potentials
which leads to larger binding energies in nuclear few-body systems.

A new type of interaction, which respects the local behavior of
traditional NN interactions at longer ranges but exhibits a
non-locality at shorter distances, was recently proposed by
Doleschall \emph{et al.}~\cite{dol03:67,dol04:69}. The authors
explore the extent to which effects of multi-nucleon forces can be
absorbed by non-local terms in the NN interaction. They
investigated if it is possible to introduce non-locality in the NN
interaction so that it correctly describes the three-nucleon bound
states, while still reproducing NN scattering
data with high precision. The so called IS version of this
interaction, introduced in Ref.~\cite{dol03:67}, contains
short-range non-local potentials in $^1S_0$ and $^3S_1-^{3\!\!}D_1$
partial waves while higher partial waves are taken from Argonne
$v_{18}$. In this study we are using the IS-M version, which
includes non-local potentials also in the $P$ and $D$
waves~\cite{dol04:69}. We note that, for this
particular version, the on-shell properties of the triplet $P$-wave
interactions have been modified in order to improve the description
of $3N$ analyzing powers. 
Unfortunately, this gives a
slightly worse fit to the Nijmegen $^{3\!}P$ phase shifts.

\begin{table}[t]
  \caption{Point-proton ($r_p$) and point-neutron ($r_n$) rms radii 
and binding energies ($E_{\rm B}$) of $^{4,6,8}$He isotopes.
The calculated values were obtained within the {\it ab initio} NCSM.
The experimental values are from Refs. 
\protect\cite{4He_chrad,6He_chrad,Tanihata,Alkhazov,Shostak,A=5-7,A=8}.
  \label{tab:r_p}}
  \begin{ruledtabular}
    \begin{tabular}{cccc}
$r_p$ [fm]  & Expt.    & CD-Bonn 2000 & INOY \\
\hline
$^4$He & 1.455(1) & 1.45(1)      & 1.37(1) \\ 
$^6$He & 1.912(18)& 1.89(4)      & 1.76(3) \\
$^8$He &          & 1.88(6)      & 1.74(6) \\
\hline
$r_n$ [fm]  & Expt.    & CD-Bonn 2000 & INOY \\
\hline
$^6$He & 2.59-2.85&  2.67(5)     &  2.55(10)\\
$^8$He & 2.69(4)  &  2.80(10)    &  2.60(10)\\
\hline
$E_{\rm B}$ [MeV] & Expt.  & CD-Bonn 2000 & INOY \\
\hline
$^4$He    & 28.296   & 26.16(6)     & 29.10(5) \\
$^6$He    & 29.269   & 26.9(3)      & 29.38(10)\\
$^8$He    & 31.408(7)& 26.0(4)      & 30.30(30)\\
    \end{tabular}
  \end{ruledtabular}
\end{table}

We performed $^4$He calculations both in the Slater determinant basis
using the Antoine code and model spaces up to $N_{\rm max}=22$ 
within the two-body effective interaction approximation and the
Jacobi-coordinate HO basis using the Manyeff code \cite{NKB00} with model spaces
up to $N_{\rm max}=20$ within either the two-body effective interaction approximation 
or the three-body effective interaction approximation. The ground-state energy 
convergence is good for both NN potentials. For the CD-Bonn 2000,
this can be seen in Fig.~1 of Ref.~\cite{ENAM04}. Our $^4$He binding energy 
and point-proton root-mean-square (rms) radii results are summarized 
in Table~\ref{tab:r_p}. We note that the point-proton rms radius is related to 
the proton charge rms radius as \cite{6He_chrad}
$\langle r_p^2\rangle = \langle r_c^2 \rangle - \langle R_p^2\rangle 
- \langle R_n^2\rangle (N/Z)$, with $(\langle R_p^2\rangle)^{1/2}=0.895(18)$ fm 
\cite{R_p}, the charge radius of the proton and $\langle R_n^2\rangle=-0.120(5)$ fm$^2$
\cite{R_n}, the mean-square-charge radius of the neutron. We observe that the
CD-Bonn 2000 underbinds $^4$He by about 2 MeV, but describes the point-proton rms radius
in agreement with experiment. The INOY NN potential, on the other hand, overbinds
$^4$He by 800 keV and underestimates the point-proton rms radius. We note that our INOY $^4$He
results are in perfect agreement with those obtained by the Faddeev-Yakubovski calculations
of Ref.~\cite{Lazauskas}. 

Our calculations for $^6$He and $^8$He nuclei were performed in model spaces
up to $N_{\rm max}=16$ and $N_{\rm max}=12$, respectively, for a wide range 
of HO frequencies. 

\begin{figure}[t]
  \includegraphics*[width=0.55\columnwidth,angle=90]{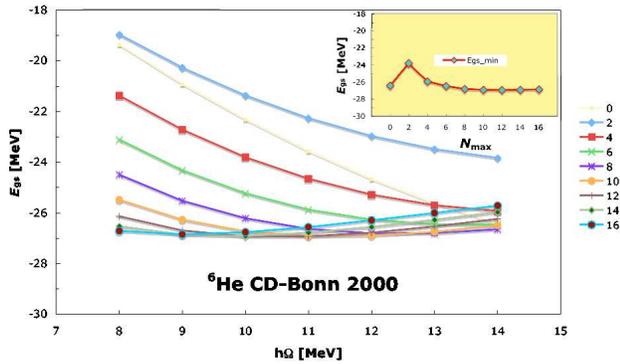}
  \caption{The $^6$He ground-state energy dependence on the HO frequency
for different model-spaces sizes from $N_{\rm max}=0$ to $N_{\rm max}=16$
obtained using the CD-Bonn 2000 NN potential. The inset demonstrates how 
the values at the minima of each curve converge with increasing $N_{\rm max}$. 
  \label{He6_cdb_gs}}
\end{figure}
\begin{figure}[hbtp]
  \includegraphics*[width=0.55\columnwidth,angle=90]{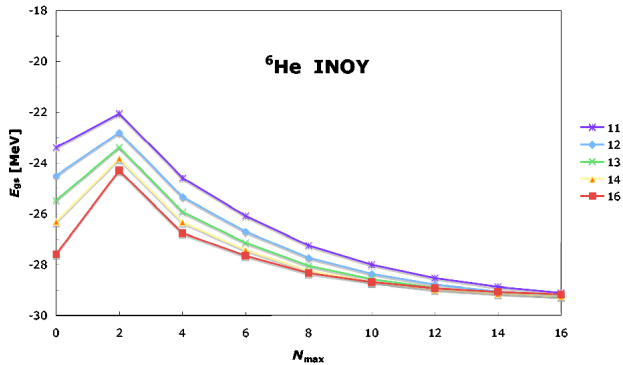}
  \caption{The $^6$He ground-state energy dependence on the model space size
for different HO frequencies from $\hbar\Omega=11$ MeV to $\hbar\Omega=16$ MeV
obtained using the INOY NN potential. 
  \label{He6_inoy_gs_nmax}}
\end{figure}

The $^6$He ground-state energy dependence on the HO frequency for different model spaces
is shown in Fig.~\ref{He6_cdb_gs} for the CD-Bonn 2000. In Fig.~\ref{He6_inoy_gs_nmax},
we show the $^6$He ground-state energy dependence on the model-space size for different 
HO frequencies obtained using the INOY NN potential.
We observe a quite different convergence trend for the two potentials.
For the INOY, the convergence is very uniform with respect to the HO frequency with systematic
changes with $N_{\rm max}$. The convergence with increasing $N_{\rm max}$ is evident. 
We extrapolate, e.g. assuming an exponential dependence on $N_{\rm max}$ 
as $E(N_{\rm max})=E_\infty+a \;{\rm exp}(-bN_{\rm max})$, that the converged
INOY ground-state energy will slightly overbind $^6$He. The ground-state energy
convergence for the CD-Bonn 2000
is quite different with a stronger dependence on the frequency, with minima shifting to lower
frequency with basis size increase, and an overall weaker dependence on $N_{\rm max}$ as seen 
in the inset of Fig.~\ref{He6_cdb_gs}.
Contrary to the INOY, the CD-Bonn underbinds $^6$He by more than 2 MeV, which is typical for the
standard high-precision NN potentials \cite{GFMC_04}.

\begin{figure}[t]
  \includegraphics*[width=0.55\columnwidth,angle=90]{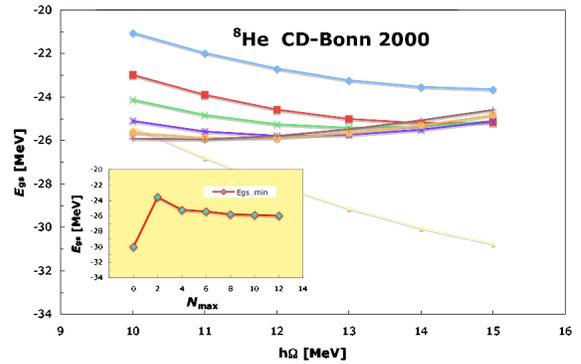}
  \caption{The same as in Fig.~\protect\ref{He6_cdb_gs}, but for $^8$He and
model-spaces from $N_{\rm max}=0$ to $N_{\rm max}=12$.
  \label{He8_cdb_gs}}
\end{figure}
\begin{figure}[hbtp]
  \includegraphics*[width=0.55\columnwidth,angle=90]{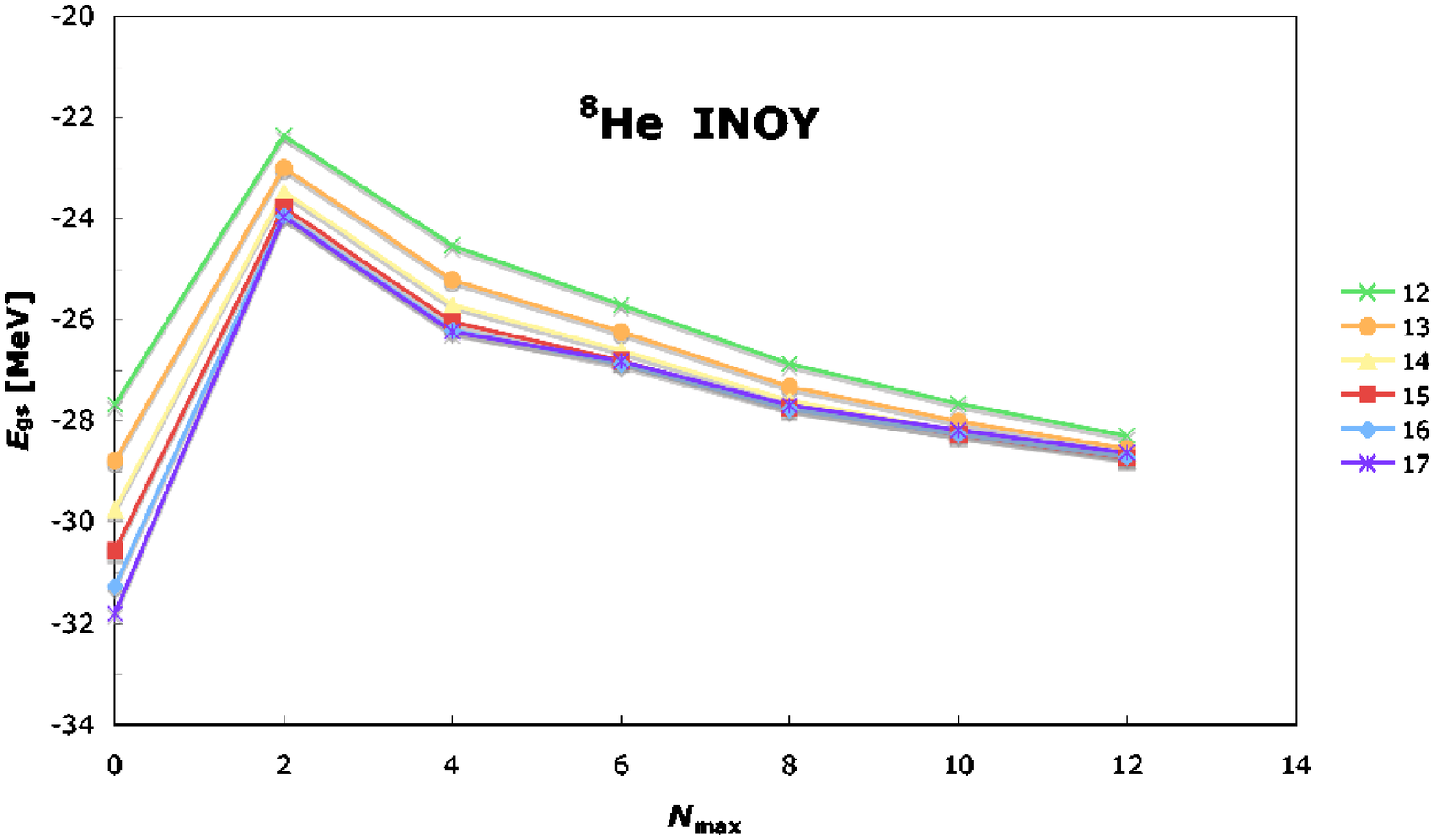}
  \caption{The same as in Fig.~\protect\ref{He6_inoy_gs_nmax}, but for $^8$He
and HO frequencies from $\hbar\Omega=12$ MeV to $\hbar\Omega=17$ MeV.
  \label{He8_inoy_gs_nmax}}
\end{figure}

The same ground-state energy dependencies for $^8$He are shown in Figs.~\ref{He8_cdb_gs}
and \ref{He8_inoy_gs_nmax}. Here, the INOY extrapolation is more difficult as, due to the
complexity of the calculations, we are limited to model spaces up to $N_{\rm max}=12$.
Our binding energy results are summarized in Table \ref{tab:r_p}. The CD-Bonn 2000 
and the INOY NN potentials underbind $^8$He by about 5 MeV and 1 MeV, respectively.
Our calculation suggest that the CD-Bonn 2000 predicts $^6$He bound but $^8$He unbound.
The INOY predicts both $^6$He and $^8$He bound. The isospin dependence of the binding 
energies is wrong for the CD-Bonn. A very similar situation was found for the Argonne 
NN potentials in Ref. \cite{GFMC_04}. Those NN potentials, at the same time, 
predict also the $^6$He unbound \cite{GFMC_04}.
The CD-Bonn NN potential must be augmented by three-nucleon
interaction to achieve a correct description of binding energies. The INOY NN potential
improves on the isospin dependence of binding energies. As this potential absorbs some
three-nucleon effects in its nonlocal part, it supports the expectation that a three-nucleon
interaction should improve the isospin dependence of binding energies. At the same time,
a three-nucleon interaction can hardly be added to the INOY NN potential as it was already 
fine-tuned to reproduce $A=3$ binding energies. Therefore, it is difficult to see, how
to correct its still not quite right binding energy predictions for the He isotopes. 

\begin{figure}[t]
  \includegraphics*[width=0.55\columnwidth,angle=90]{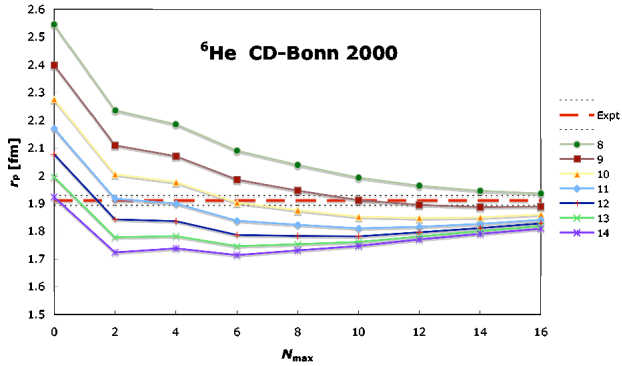}
  \caption{The $^6$He point-proton rms radius dependence on the model space size
for different HO frequencies from $\hbar\Omega=8$ MeV to $\hbar\Omega=14$ MeV
obtained using the CD-Bonn 2000 NN potential. 
The experimental value is from Ref. \protect\cite{6He_chrad}.
  \label{He6_cdb_rp}}
\end{figure}
\begin{figure}[hbtp]
  \includegraphics*[width=0.55\columnwidth,angle=90]{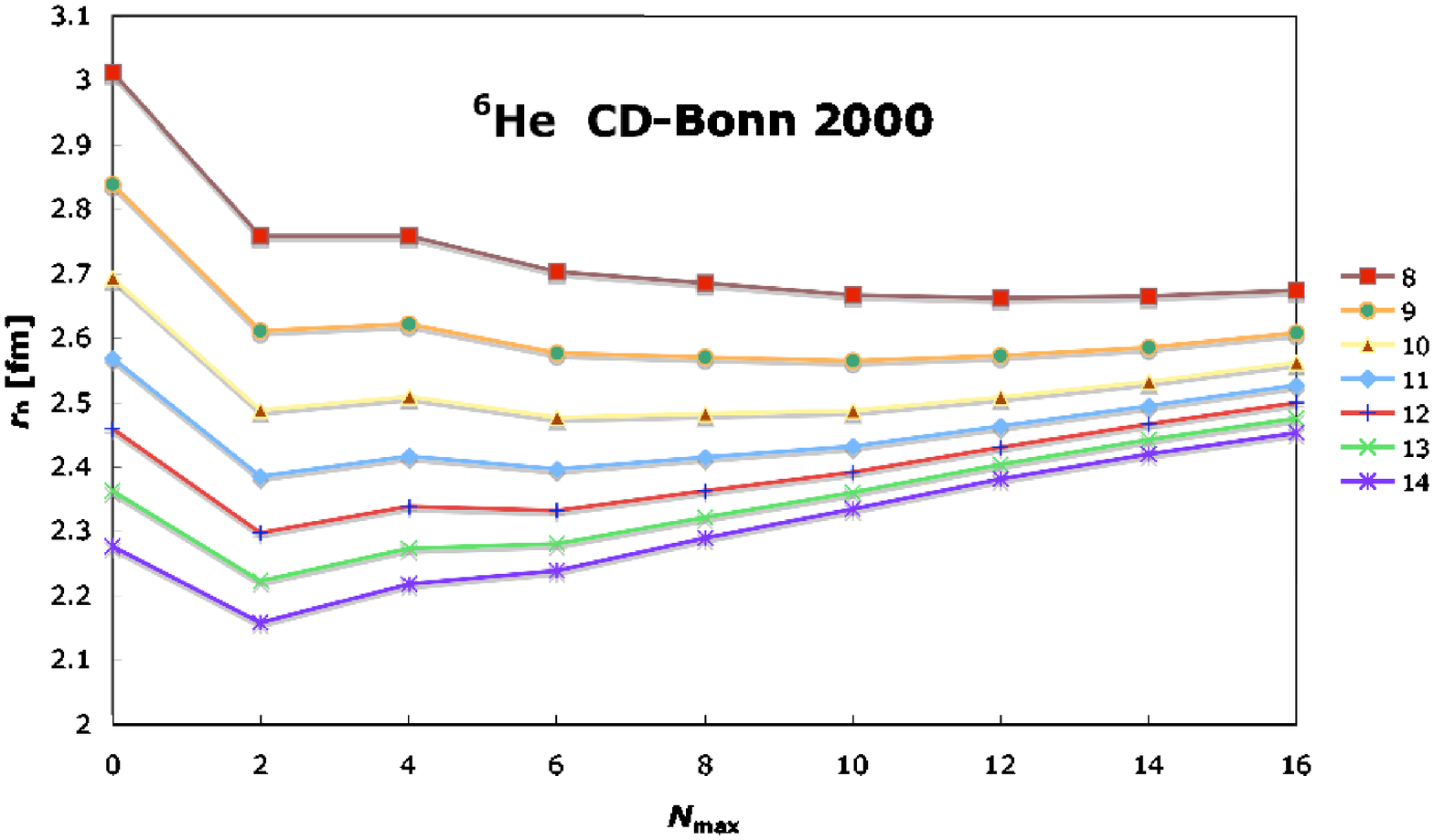}
  \caption{The $^6$He point-neutron rms radius dependence on the model space size
for different HO frequencies from $\hbar\Omega=8$ MeV to $\hbar\Omega=14$ MeV
obtained using the CD-Bonn 2000 NN potential. 
  \label{He6_cdb_rn}}
\end{figure}
\begin{figure}[t]
  \includegraphics*[width=0.55\columnwidth,angle=90]{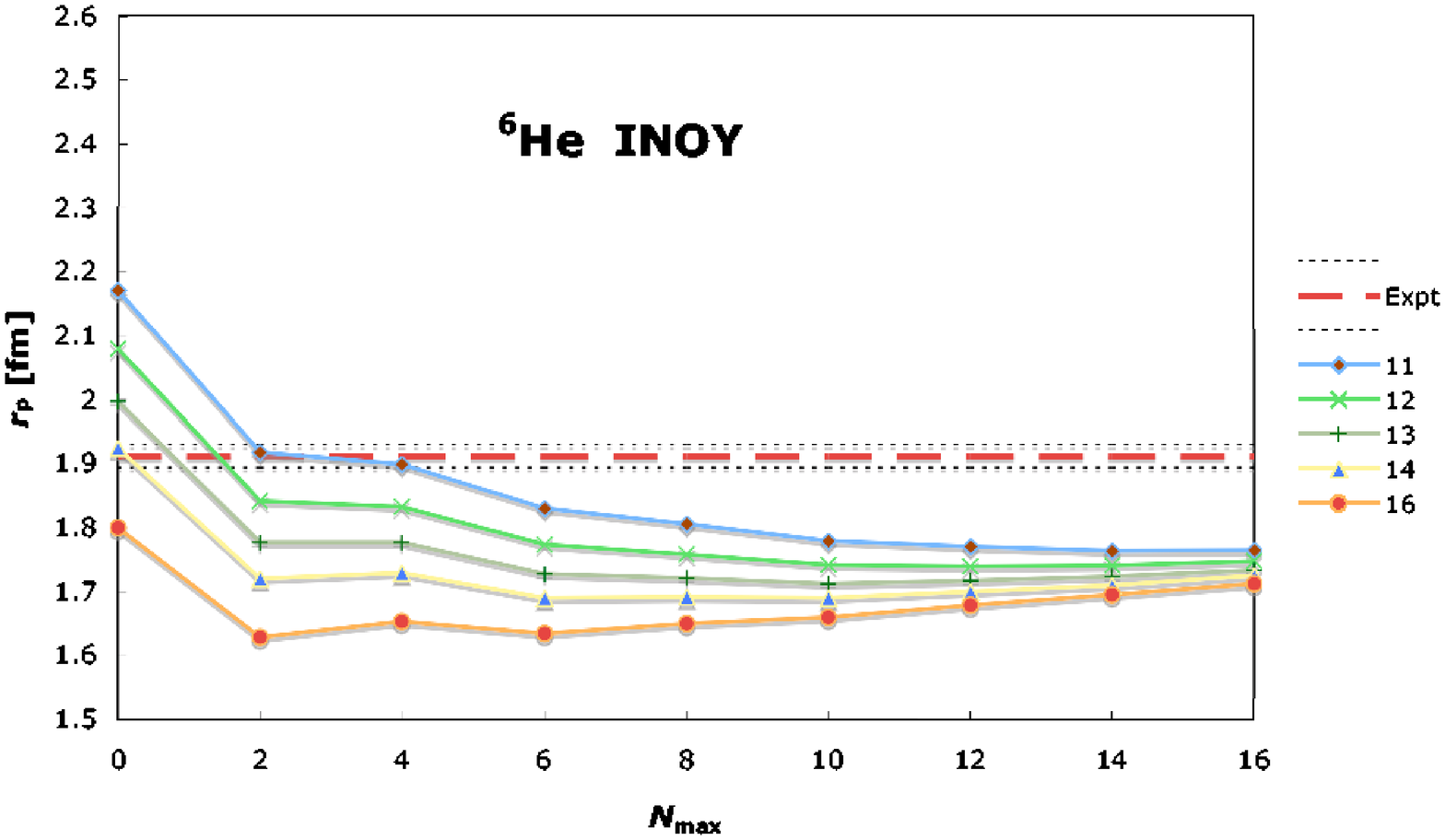}
  \caption{The same as in Fig.~\protect\ref{He6_cdb_rp}, but for the INOY NN potential
and HO frequencies from $\hbar\Omega=11$ MeV to $\hbar\Omega=16$ MeV.
  \label{He6_inoy_rp}}
\end{figure}
\begin{figure}[hbtp]
  \includegraphics*[width=0.55\columnwidth,angle=90]{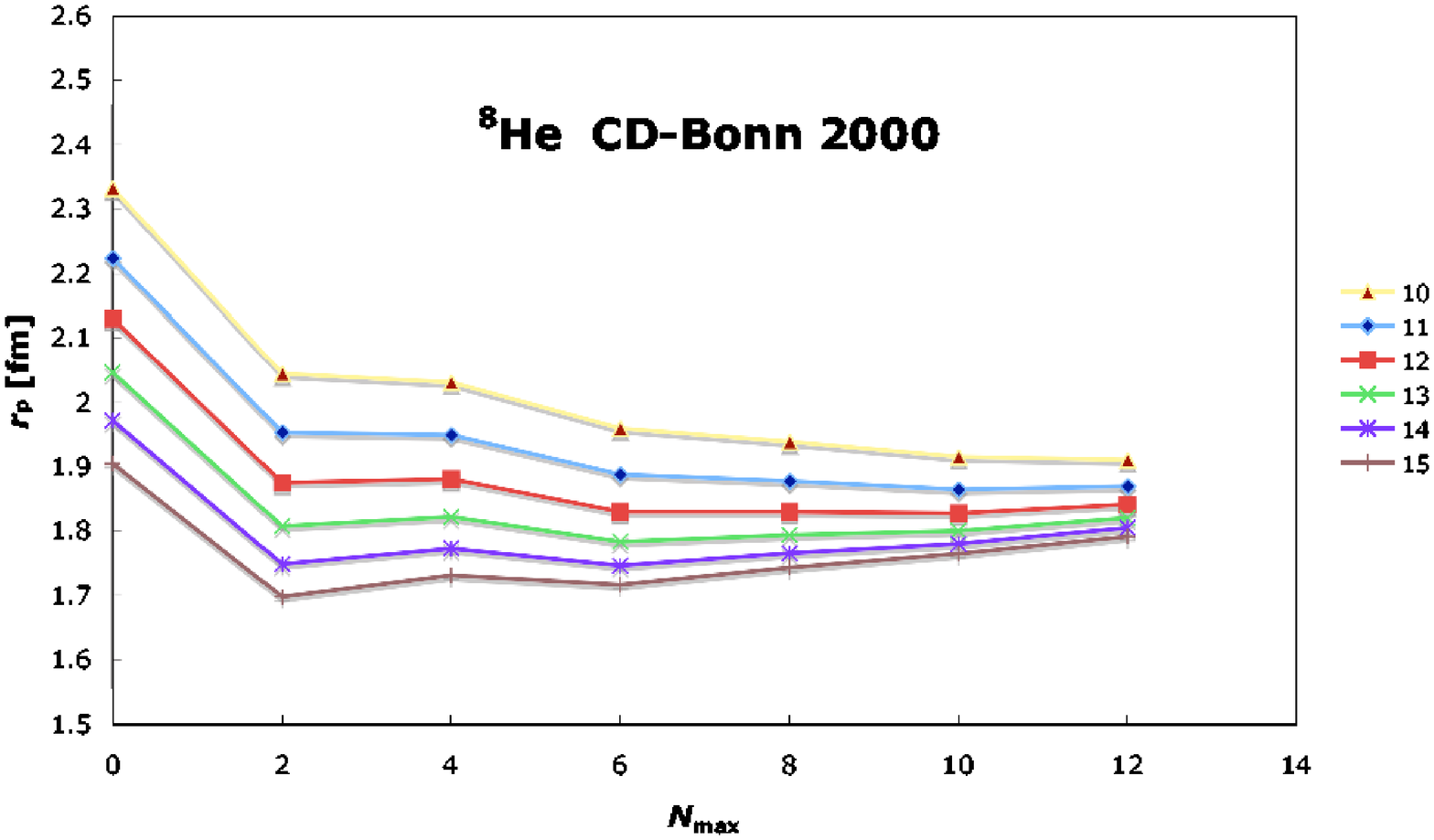}
  \caption{The $^8$He point-proton rms radius dependence on the model space size
for different HO frequencies from $\hbar\Omega=10$ MeV to $\hbar\Omega=15$ MeV
obtained using the CD-Bonn 2000 NN potential. 
  \label{He8_cdb_rp}}
\end{figure}

\begin{figure}[t]
  \includegraphics*[width=0.55\columnwidth,angle=90]{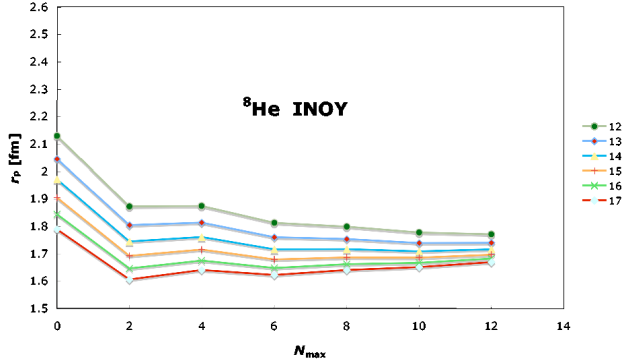}
  \caption{The same as in Fig.~\protect\ref{He8_cdb_rp}, but for the INOY NN potential
and HO frequencies from $\hbar\Omega=12$ MeV to $\hbar\Omega=17$ MeV.
  \label{He8_inoy_rp}}
\end{figure}

Our point-nucleon rms results are presented in 
Figs.~\ref{He6_cdb_rp}-
\ref{He8_inoy_rp} and summarized 
in Table~\ref{tab:r_p}. In the figures, we show the model-space size dependence 
of the rms radii for different HO frequencies. A general feature is a decrease
of the HO frequency dependence with increasing model-space size defined by $N_{\rm max}$. 
In all cases, the rms radii exhibit convergence. The $^6$He point-proton rms
radius experimental value is shown as a dashed line in Figs.~\ref{He6_cdb_rp} 
and \ref{He6_inoy_rp} with the dotted lines indicating the experimental error.
The CD-Bonn 2000 $^6$He point-proton rms radius, Fig.~\ref{He6_cdb_rp}, stabilizes
at $N_{\rm max}=16$ for the HO frequencies of $\hbar\Omega=9$ and 10 MeV, while
it is still decreasing for $\hbar\Omega=8$ MeV and it is
increasing for the HO frequencies higher than $\hbar\Omega=10$ MeV. Clearly,
the stable result is very close to the experimental value. We estimate the error
of our calculation at $N_{\rm max}=16$ from the HO frequency dependence.
We note that we published the $^6$He CD-Bonn point-proton rms radii in Ref.~\cite{NCSM6}.
Those results were obtained using the HO frequency of $\hbar\Omega=13$ MeV in 
$N_{\rm max}=6,8$ and 10 model spaces. Our $N_{\rm max}=10$ value, 1.763 fm, was then
compared to experiment in Ref.~\cite{6He_chrad}. We can see from Fig.~\ref{He6_cdb_rp}
that the radius is still increasing with $N_{\rm max}$ for that frequency and reaches, 
e.g. 1.819 fm at $N_{\rm max}=16$. From our present results obtained up to $N_{\rm max}=16$ 
for a wide range of HO frequencies we arrive at the CD-Bonn 2000 point-proton rms radius
of 1.89(4) fm that, taken into account the error bars, agrees with the experimental 
value of 1.912(18) fm. The point-neutron rms radius shows a stronger dependence on the HO
frequency and a slower convergence as seen in Fig.~\ref{He6_cdb_rn}. 
This is to be expected as the neutron halo is extended and a large HO basis is needed 
to describe it properly. Nevertheless, we observe a reasonable stability of the neutron
rms radius at lower HO frequencies that allows us to estimate its
CD-Bonn 2000 value to be 2.67(5) fm.  


%

We observe a better convergence for the INOY NN potential not only for the binding
energies but also for the radii. This is apparent from Fig.~\ref{He6_inoy_rp}. For
this NN potential, we find the $^6$He point-proton rms radius to be 1.76(3) fm. This
is significantly less than in experiment. Clearly, the INOY NN potential underpredicts
both the $^4$He and $^6$He point-proton rms radii.  

Our $^8$He point-proton rms radius results are shown in Figs.~\ref{He8_cdb_rp} 
and \ref{He8_inoy_rp} for the CD-Bonn 2000 and INOY potentials, respectively.
Based on the basis size and the HO frequency dependence, we predict the 
$^8$He point-proton rms radius to be 1.88(6) fm based on our CD-Bonn results.
The INOY NN potential gives a smaller value, 1.74(6) fm, consistently with the 
smaller $^4$He and $^6$He results. In both cases, the $^8$He point-proton radius is 
slightly smaller then the corresponding one in $^6$He. Taking into account the 
uncertainties, however, the differences are insignificant.

In conclusion, we performed large-scale {\it ab initio} NCSM calculations
for $^4$He, $^6$He and $^8$He isotopes. We used the high-precision CD-Bonn 2000
and the INOY NN potentials and obtained results for binding energies and point-nucleon
rms radii. Using the CD-Bonn 2000, we obtained the point-proton rms radii of $^4$He
and $^6$He in agreement with experiment and predict the $^8$He point-proton rms radius
to be 1.88(6) fm. The INOY NN potential, on the other hand, underestimates both
$^4$He and $^6$He experimental point-proton rms radii. The CD-Bonn 2000 underbinds
the He isotopes as is typical for the standard high-precision NN interactions.
It must be augmented by a three-nucleon interaction. It is conceivable that this
can be done in a way that will not change the charge radii. The INOY NN potential
gives binding energies closer to experiment. However, it is not obvious, how
the charge radii results can be brought to agreement with experiment when using 
this potential. It can hardly be augmented by a three-nucleon interaction 
as it was already fine-tuned to describe the $A=3$ system. 

This work was partly performed under the auspices of the
U. S. Department of Energy by the University of California, Lawrence
Livermore National Laboratory under contract No. W-7405-Eng-48. Support
from the LDRD contract No.~04--ERD--058, and from 
U.S. DOE, OS (Work Proposal Number SCW0498) is acknowledged.

\end{document}